\documentclass[9pt,twocolumn,twoside]{osajnl}
\usepackage{amssymb}
\usepackage{verbatim}
\usepackage{xcolor}
\journal{ol} 

\setboolean{shortarticle}{true}

\title{Chirp-compensated pulsed titanium-sapphire laser system for precision spectroscopy}

\author[1]{J. Hussels}
\author[2]{C. Cheng}
\author[1]{E.J. Salumbides}
\author[1,*]{W. Ubachs}

\affil[1]{Department of Physics and Astronomy, LaserLaB, Vrije Universiteit, De Boelelaan 1081, 1081 HV Amsterdam, The Netherlands}
\affil[2]{Hefei National Laboratory for Physical Sciences at Microscale,
University of Science and Technology of China, Hefei, 230026 China}

\affil[*]{Corresponding author: w.m.g.ubachs@vu.nl}

\doi{\url{http://doi.org/10.1364/OL.401703}}

\begin{abstract}
Active frequency-chirp compensation for a narrowband pulsed  Titanium-Sapphire laser system is demonstrated using an intra-cavity electro-optic modulator resulting in improved spectral resolution and stability.
With referencing to an optical frequency comb and further residual frequency chirp detection from shot-to-shot measurements, the resulting laser pulses are frequency up-converted for high-precision spectroscopy measurements in the VUV regime, where the relative uncertainty contribution due to frequency chirp is pushed to the $5 \times 10^{-11}$ level.
\end{abstract}

\setboolean{displaycopyright}{true}

\begin{document}

\maketitle


Frequency metrology experiments are generally performed with continuous-wave lasers, allowing for long term stability and highly accurate read-out of the center-frequency of narrow resonances, such as in optical clocks~\cite{Rosenband2008,Bloom2014}. However, in a variety of applications the use of pulsed lasers in precision experiments is required, for the necessity of power density, for triggered detection, in cases of multi-photon experiments, or at wavelengths in otherwise inaccessible regions. In some fundamental physics experiments lasers with a well-controllable pulse structure are prerequisite, such as in investigations of positronium~\cite{Fee1991}, muonium, anti-protonic helium~\cite{Hori2011}, and muonic hydrogen. Moreover, laser pulses are required to reach the domain of the extreme ultraviolet, e.g. for the excitation of the He resonance at 58 nm~\cite{Eikema1996}. Time-dependent gain in the lasers and amplifiers, as well as mode-pulling phenomena inside a laser cavity during build-up of the population inversion give rise to a non Fourier-transform limited pulse and to a time dependence of its central frequency, thus causing overall offsets in the frequency measurements. Such chirp effects have been investigated over the years in traveling-wave pulsed dye amplifiers ~\cite{Fee1992,Melikechi1994,Reinhard1996,Eikema1997}, in optical parametric oscillators~\cite{White2004}, in an alexandrite laser~\cite{Bakule2000}, as well as in pulsed titanium-sapphire lasers~\cite{Hannemann2007b,Hori2009,Sprecher2013}.

\begin{figure}[ht]
\centering
\fbox{\includegraphics[width=0.9\linewidth]{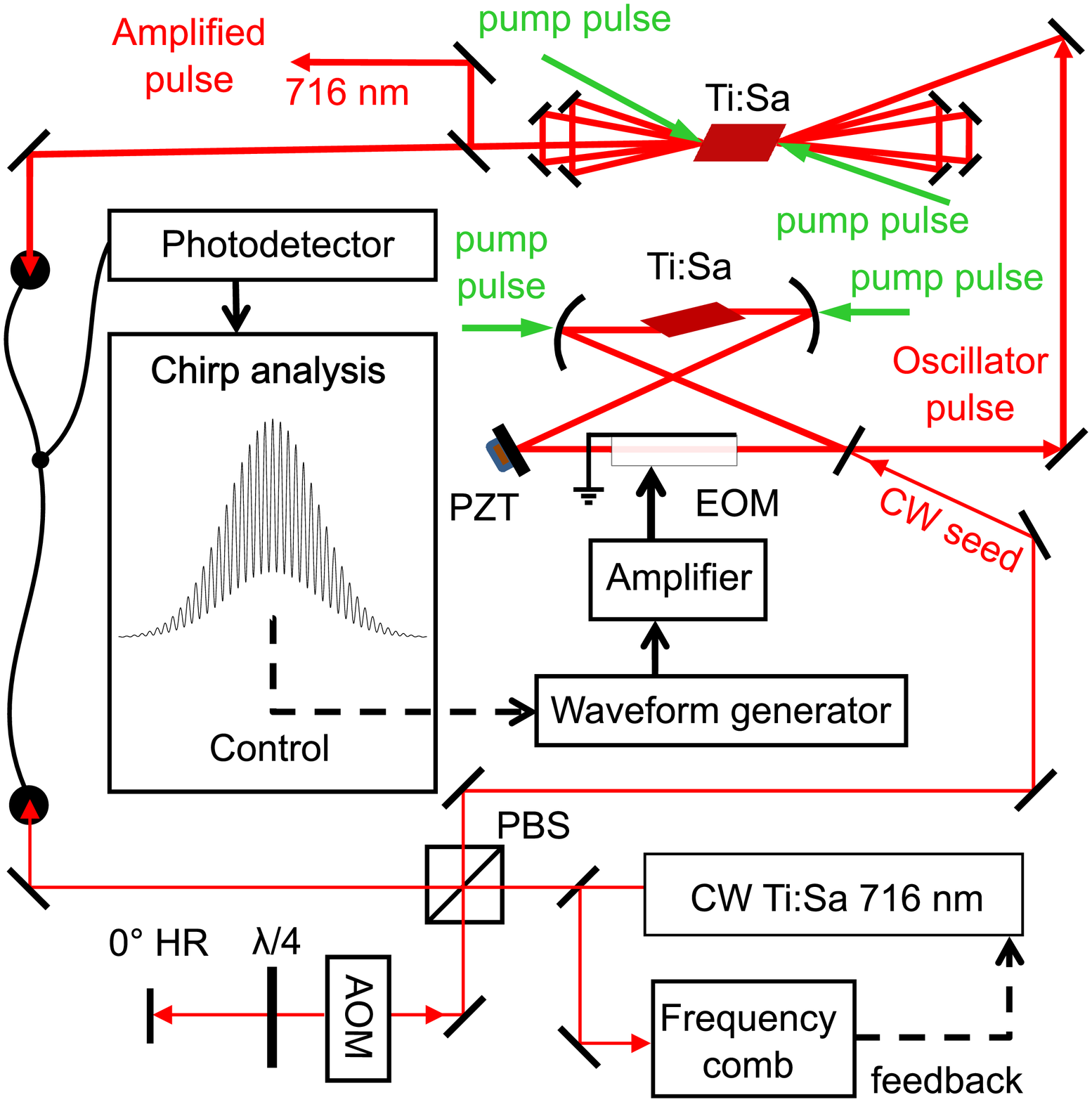}}
\caption{\small{Overview of the setup. For explanation see main text.}}
\label{fig:setup}
\end{figure}

Here we report studies on chirp reduction, measurement, and compensation in nanosecond pulsed Ti:Sa lasers, by extending our passive mode-locking scheme~\cite{Hannemann2007b} to an active scheme following a method explored by Hori and Dax~\cite{Hori2009}. The aim is to improve the accuracy limits and operate the Ti:Sa laser away from the peak of the gain profile at a wavelength of 716 nm.
The chosen wavelength allows for comparing the resulting absolute frequency uncertainty of actively chirp-compensated and frequency upconverted pulses to a molecular resonance in H$_2$.

The optical layout of the Titanium:Sapphire (Ti:Sa) laser system, shown in Fig.~\ref{fig:setup}, consists of two main components: an oscillator cavity, in which pulses of about 80~ns duration are generated, and a bowtie multipass traveling-wave amplifier, where the cavity output is further enhanced for subsequent efficient harmonic upconversion.
The oscillator is built in the form of a stable ring cavity with a Brewster-cut Ti:Sa crystal inserted as the gain material.
A Q-switched Nd:YAG pump laser operating at 10 Hz repetition rate produces pulses of 300~mJ at 532~nm, where the major part is used for the amplifier and a fraction of 10~mJ is split for pumping the oscillator.
In the oscillator, the pump pulses are propagated through dichroic cavity mirrors and focused on the crystal from both sides to earn maximum volume utilization.  Measurements with both an unseeded and an injection-seeded Nd:YAG laser were performed and no significant differences found.

The output of a cw Ti:Sa ring laser (Coherent MBR E-110) at a power of 200~mW is injected into the cavity as a seed.
To suppress broadband super-fluorescent pulses produced by the Ti:Sa crystal at wavelengths of higher gain, the reflectivity of the cavity mirrors is designed to be 99.9\% at 716~nm and less than 40\% for wavelengths longer than 740~nm.
The output coupler has a reflectivity of 98\% (equivalent cavity finesse of about 300) at the desired wavelength (716~nm).
Pulse energies of up to 1~mJ can be generated from the Ti:Sa oscillator and sent to the multi-pass
amplifier. After 10 amplification steps through the crystal inside the amplifier, also pumped from both sides, the pulse energies are enhanced to about 40~mJ.

In order to generate controlled and stable pulses, the oscillator cavity is stabilized by locking to the seed light using a H\"{a}nsch-Couillaud (HC) locking method~\cite{Hansch1980}. The locking system has a bandwidth of about 10~kHz and the feedback signal is sent to a fast piezoelectric ceramics (PZT) attached to one of the cavity mirrors.
The frequency accuracy of the seed light is guaranteed by locking the cw Ti:Sa laser to an optical frequency comb referenced to a Cs atomic clock.
The short-term (one second) frequency stability is a few tens of kHz and the long-term relative accuracy of 10$^{-12}$ is limited by the frequency comb.
An acousto-optic modulator (AOM) is implemented in a double-pass scheme for reasons of scanning the cw laser frequency for spectroscopic acquisition and for generating a large enough frequency shift for the chirp measurements.

When the Ti:Sa crystal inside the cavity is optically pumped, the refractive index of the crystal changes due to population inversion. This results in a fast change in optical path length inside the cavity, and thus the frequency of the generated pulses will undergo cavity mode pulling. The H\"{a}nsch-Couillaud lock is too slow to follow the cavity mode excursion during the build-up of gain. During the pulse evolution the excited state population of the Ti:Sa decays, emitting 716 nm light, thus resulting in a continuous, and in first order linear change in the instantaneous frequency during the pulse, \emph{i.e.} chirp. The settings of the H\"{a}nsch-Couillaud lock as well as the pump power can be manipulated to affect the pulling and chirp effects in the cavity, as was demonstrated in the 780-860 nm optimum gain region of Ti:Sa~\cite{Hannemann2007b}.

\begin{figure}
\centering
\fbox{\includegraphics[width=\linewidth]{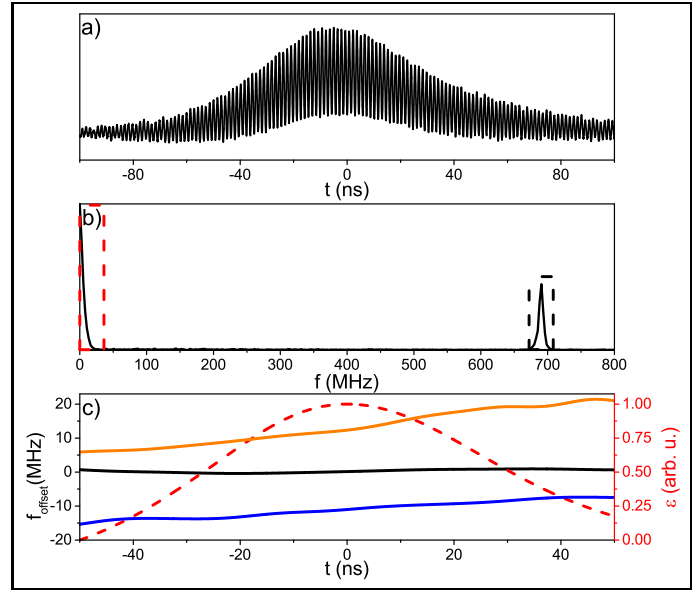}}
\caption{\small{a) Beat-note signal of a pulse combined with the unshifted cw light, displaying 69 periods of the beat over 100 ns, at an average of 58 samples/period. b) Fourier-transform of this pulse, showing the two components (see text). c) Dashed curve (red) represents the pulse envelope $\epsilon(t)$; Full curves represent the instantaneous offset without chirp compensation (blue), for unoptimized chirp compensation (orange), for optimized chirp compensation (black).}}
\label{fig:meas}
\end{figure}

The chirp of the Ti:Sa oscillator-amplifier laser is measured by combining a fraction of the output pulses with cw-laser light, after propagating both beams through a single-mode fiber for spatial mode overlap. Their beat-note as detected on a fast photo detector connected to an 8 GHz band-width oscilloscope (sampling rate 40 Gsamples/s) is shown in Fig. ~\ref{fig:meas}a. The Fourier transform of the signal shows two peaks in the frequency spectrum as plotted in Fig.~\ref{fig:meas}b, where the low frequency peak is associated with the pulse envelope. The high frequency peak, associated with the 690 MHz shift in the double pass AOM, contains information on the chirp.
Reversing the Fourier transformation separately for the two peaks generates the pulse envelope $\epsilon(t)$ and the phase evolution $\Phi(t)$ of the oscillatory function~\cite{Hannemann2007b}.

The instantaneous frequency offset $f_{\rm offset}(t)$ at time $t$ is obtained via:
\begin{equation}
    f_{\rm offset}(t) = \frac{1}{2\pi}\frac{d\Phi}{dt}-f_{\rm AOM}
\end{equation}
where any offset from 0 is due to the pulling and chirp effects. This $f_{\rm offset}(t)$ is displayed in Fig.~\ref{fig:meas}c for three specific cases (to be discussed below).  The average offset is defined by
\begin{equation}
    <f_{\rm offset}> = \int f_{\rm offset}(t)w(t)dt
    \label{off}
\end{equation}
with the pulse envelope $\epsilon(t)$ taken as a weight factor $w(t)= \epsilon(t) / (\int \epsilon(t)dt)$.
The average slope of the chirp profile is defined by
\begin{equation}
    <f_{\rm slope}> = \int \frac{df_{\rm offset}}{dt}(t)w(t)dt
    \label{slope}
\end{equation}

To manipulate and actively compensate the chirp effects, an Electro-Optic Modulator (EOM), made of a LiTaO$_{3}$-crystal, is placed inside the oscillator cavity.
The optical end faces of the EOM are coated to obtain a transmission of 99.7\% to minimize optical losses.
The EOM is coated with a gold layer on the side faces, where the lower side is grounded and the upper side is connected to an electric amplifier, which is driven by an arbitrary waveform generator (AWG).
For one pass through the EOM, the acquired phase is:
\begin{equation}
    \Phi(t)=-\frac{\pi n_0^3 L}{\lambda d}\left[ \left( \frac{n_e}{n_0} \right)^3 \gamma_{33}-\gamma_{13}\right]U(t)
    \label{phi}
\end{equation}
where $n_o=2.175$ and $n_e=2.180$ are ordinary and extraordinary refractive indices of the EOM of length $L=16$ mm and thickness $d=2.7$ mm,
and $\gamma_{33}=33 \times 10^{-12}$ m/V and $\gamma_{13}=8 \times 10^{-12}$ m/V the relevant electro-optical constants.
This results in a phase shift of 0.017 radians per Volt per pass. The wavelength is set at $\lambda=716$ nm and $U$ is the applied voltage.

The total frequency shift is produced by the cumulative effect of multiple cavity-roundtrips, involving mode-pulling of the cavity, chirp in the Ti:Sa crystal, counteracted by the phase effects imposed by the EOM.
This bears the advantage that low voltages of 30-35 V are already sufficient to change the instantaneous frequency significantly.
The chirp in the Ti:Sa amplifier is relatively small~\cite{Hannemann2007b}, in view of the limited number of passes (10 times), and the fact that the population inversion is produced some 200 ns before arrival of the oscillator pulse. During the amplification process this inversion, and hence the refractive index of the amplifier crystal remain constant, only decaying on the excited state time scale ($3 \mu$s). It is noted that all chirp measurements pertain to the combined oscillator-amplifier system.

The EOM is driven by an AWG, programmed to provide a sawtooth-like triangle function, starting with a linear drop of -34 V (after electrical amplification) during the initial 500 ns, followed by an instantaneous return to 0 V.
The limited slew rate (2000V/$\mu$s) of the amplifier delivers an exponentially increasing voltage over the EOM during return, resulting in an effective voltage profile as shown in Fig. ~\ref{fig:meas}b. This profile, imposed on the EOM, is well-suited to counteract the disturbances of the oscillator cavity.
The linear decay in the initial phase affects the average frequency, therewith canceling the mode pulling effect and stabilizing the oscillator.
The exponential return in the latter part is exploited to produce a fast voltage change over the EOM, therewith effectively canceling chirp effects. Different functions have been used, but the fast exponential rise proved to be the most effective.

\begin{figure}[htbp]
\centering
\fbox{\includegraphics[width=\linewidth]{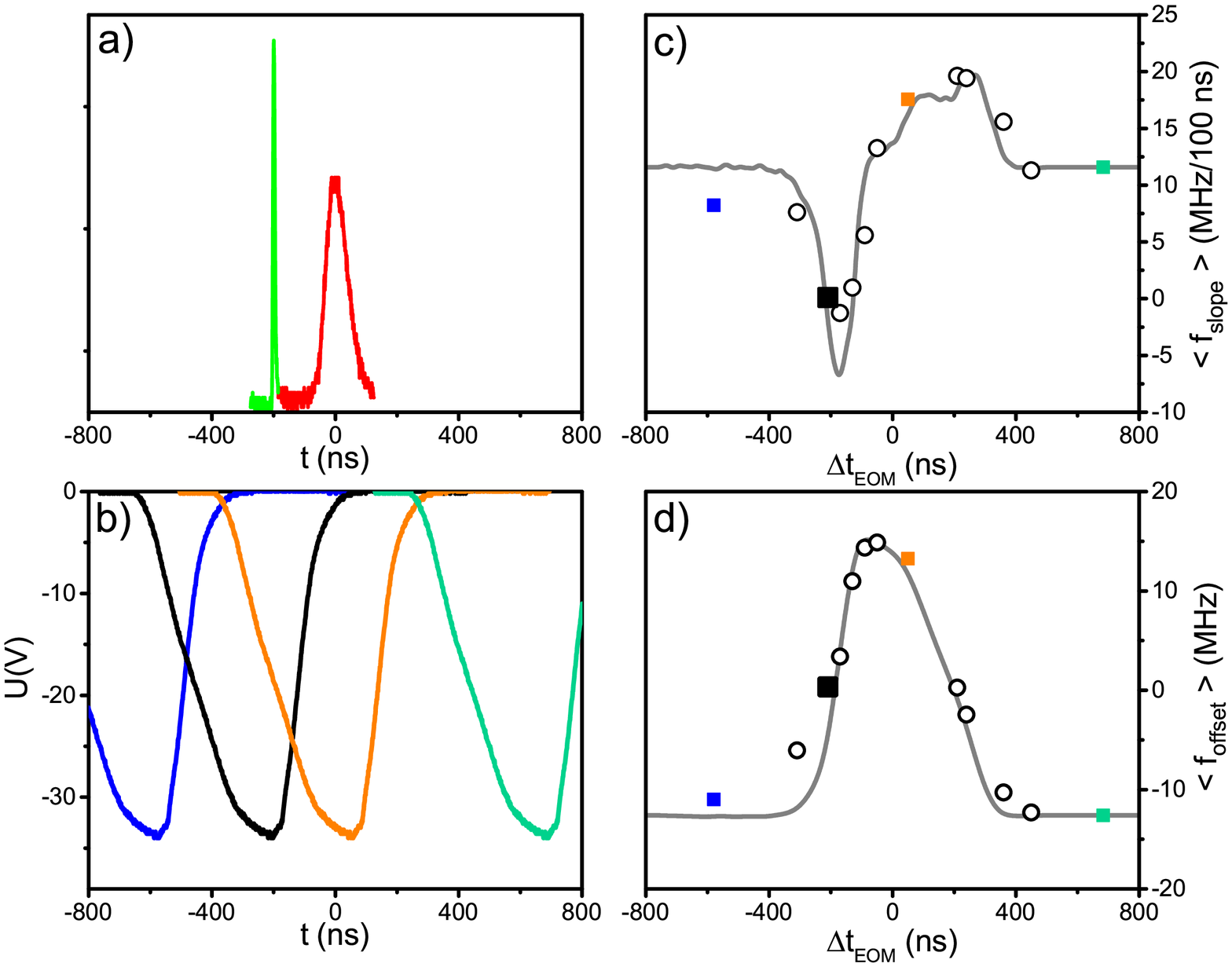}}
\caption{\small{a) The Nd:YAG pump pulse followed by - a slightly asymmetric - Ti:Sa pulse, at a build-up time of 200 ns. The amplitude of the two pulses are not to scale with respect to eachother.  b) Measured voltage evolution over the EOM at different time settings. c) Average slope of the chirp profile, as a function of the delay of the voltage pattern over the EOM. d) Average offset of the chirp profile, as a function of the delay of the voltage pattern over the EOM. The (square) colour coded data points in c) and d) correspond to the voltages settings in b) and the color-coded chirp profiles in Fig.~\ref{fig:meas}c). The black squares corresponds to the setting of optimized anti-chirp. The curves are simulated, the voltage pattern of b) is used as input for the simulations (see text).}}
\label{fig:EOMV}
\end{figure}

The timing of the voltage pattern, U(t), with respect to the Ti:Sa pulse, $(\Delta t_{EOM})$, proved to be the most sensitive parameter for the chirp compensation. Fig.~\ref{fig:EOMV}c and d show $<f_{\rm offset}>$ and $<f_{\rm slope}>$ of the measured chirp profile, using Eqs.~(\ref{off}) and (\ref{slope}), as a function of the timing of the voltage, where $(\Delta t_{EOM})=0$ corresponds to the situation where the dip in the voltage and the peak of the Ti:Sa pulse overlap.

In order to produce quantitative understanding of the chirp phenomena simulations were performed, resulting in the curves shown in Fig.~\ref{fig:EOMV}c and d. The build-up time shown in Fig.~\ref{fig:EOMV}a and the known round-trip time of 1.3 ns (optical path length of 39 cm) were used to determine the number of passes through the EOM for every point in the Ti:Sa pulse. Hence, steps of 1.3 ns were taken so that every point corresponds to a different number of round-trips.  The added phase per pass is calculated with Eq.~(\ref{phi}), where the input for the Voltage are obtained from the measured profiles in Fig. ~\ref{fig:EOMV}b. To determine the accumulated phase of each point over the laser pulse evolution, the added phase per pass is summed over all the passes through the EOM.
Consequently, the change in instantaneous frequency over the Ti:Sa pulse can be determined using
\begin{equation}
    f_{\rm EOM} = \frac{1}{2\pi}\frac{\Delta\Phi}{\Delta t},
    \label{fEOM}
\end{equation}
where $\Delta\Phi$ is the difference in total accumulated phase between two sequential points ($\Delta t = 1.3$ ns).
For a certain setting of the delay of the voltage pattern over the EOM, this results in a set of simulated chirp parameters, for offset and slope.
However, these simulated chirp profiles are solely due to the EOM, and cannot be directly compared to the measurements, since the measurements also include chirp effects from both the Ti:Sa crystals in oscillator and multi-pass amplifier.
For a true comparison the average offset and slope due to the Ti:Sa must be subtracted. These can be obtained from the measurements without anti-chirp (so at very high EOM-delay).
In this way the simulations of the chirp compensation effects reflect the observed data points for $<f_{\rm offset}>$ and $<f_{\rm slope}>$ as shown as the curves in Fig.~\ref{fig:EOMV}c and d. Importantly, the optimum setting of the EOM voltage pulse, for obtaining $<f_{\rm offset}> \approx 0$ and $<f_{\rm slope}> \approx 0$ is determined, for use in spectroscopic application.

For a laser spectroscopic precision measurement the laser output (at 716 nm) is frequency upconverted in two doubling stages, with BBO and KBBF crystals, leading to the generation of 179~nm radiation to drive the GK-X Q(1) transition of molecular hydrogen in a two-photon scheme. The vacuum-ultraviolet (VUV) output beam of up to 20~$\mu$J per pulse is then intersected with a pulsed molecular hydrogen beam. After retro-reflection 2+1$'$ two-photon enhanced ionization Doppler-free spectroscopy is achieved using counter-propagating VUV-beams in combination with a delayed ionization pulse~\cite{Cheng2018}. After driving the EOM the chirp is detected and analyzed, and thereupon the EOM control parameter (the timing) is manually set for optimal compensation of the chirp phenomena.
While performing a spectroscopic measurement, the EOM is fixed in this optimal condition, but nevertheless final residual chirp effects are detected for determining pulse-to-pulse variations, and to numerically correct transition frequencies for each pulse.
Due to the frequency upconversion in the multiple doubling stages and the two-photon transition scheme, any residual instantaneous frequency offset in the fundamental pulse will be multiplied by 8, thus providing a sensitive test for our chirp compensation method.

\begin{figure}[htbp]
\centering
\fbox{\includegraphics[width=0.9\linewidth]{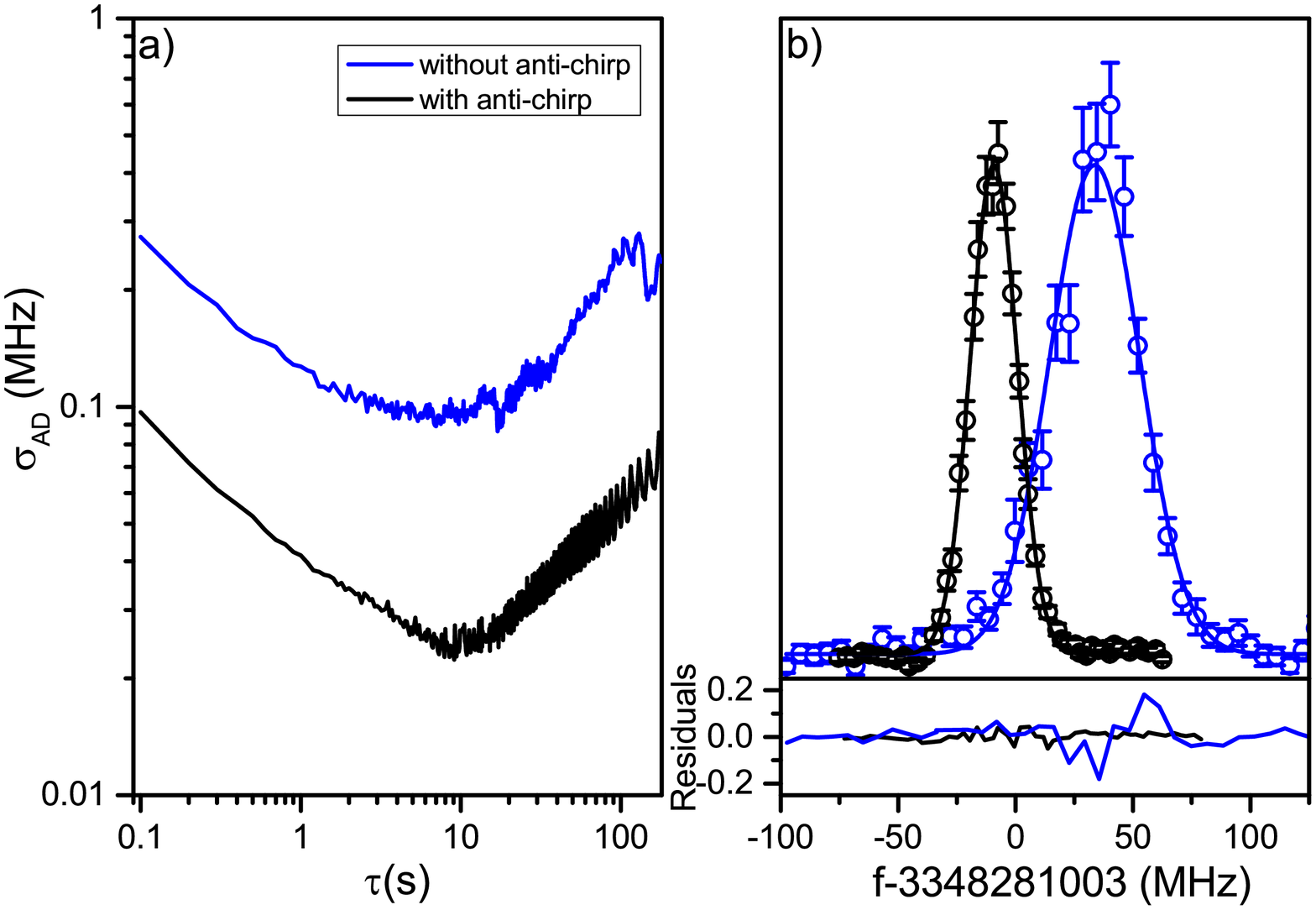}}
\caption{\small{a) Allan Deviation of the average frequency offset over 5000 sequential pulses, measured with and without compensation for the pulling and chirp effects.   b) Two-photon spectra of H$_2$ X-GK Q(1) with and without EOM-compensation and averaging over 50 shots.
The residuals show that the spectrum with active chirp compensation is Gaussian, while the passive spectrum is asymmetric.}}
\label{fig:AllanDev}
\end{figure}

Following the investigation by Hannemann et al.~\cite{Hannemann2007b}, in this experiment we have carefully chosen parameters of HC-lock setting, pump energy, and mirror reflectivities to reduce the cavity-pulling and chirp-induced frequency shifts.
While the frequency chirp was previously only passively controlled, the intracavity EOM now allows for the active frequency chirp compensation.
The highest spectral resolution for precision spectroscopy is achieved when using the longest pulse duration at zero frequency chirp, corresponding to a FT-limited spectral band-width.
At 806 nm very good control of the frequency chirp can be obtained, since near the centre of the Ti:Sa band-width changes in the index of refraction are small~\cite{Hannemann2007b}. However, the precision spectroscopy of H$_2$~\cite{Cheng2018} requires the oscillator to work at 716 nm, which is on the edge of the Ti:Sa gain profile, leading to larger frequency pulling and chirp effects,
as well as to larger pulse-to-pulse fluctuations.
The active chirp compensation with the EOM eliminates cavity frequency pulling, since by counteracting the large refractive index change the cold cavity resonance condition is preserved even in the presence of the pump pulse.
This leads to improved temporal stability of the pulses as shown by the Allan deviation of the frequency offset in Fig.~\ref{fig:AllanDev}a defined as:
\begin{equation}
    \sigma_{\rm AD}(\tau)^2 = \frac{1}{2}<(<f_{n+1}>-<f_{n}>)^2>
\end{equation}
where <f$_{n}$> is the n$^{th}$ <{f}$_{\rm offset}$> average over time $\tau$.
The Allan deviation is a factor of 3-5 lower when the active anti-chirp compensation (black curve) is engaged.
It is apparent in Fig.~\ref{fig:AllanDev}a, that 10-s averaging is the optimum, with the increase in instability at longer times probably limited by the locking electronics and pump pulse stability.

The effect of the frequency chirp compensation on the spectral resolution, which is of main importance in precision spectroscopy, is shown in Fig.~\ref{fig:AllanDev}b.
Note that the frequency (horizontal) axis in Fig.~\ref{fig:AllanDev}b includes a factor of 8 from the frequency upconversion and the two-photon excitation with respect to the fundamental Ti:Sa frequency.
The full width at half maximum (FWHM) of the Gaussian fit is reduced from 49.8 MHz without compensation to 25.9 MHz when cancelling the linear frequency chirp (<2~MHz/100~ns). This leads to a more accurate line center determination, besides counteracting the chirp. For the GK ($v=1$, $J=1$) state a lifetime has been reported of  16.9(5) ns~\cite{Hoelsch2018}, corresponding to a natural line-width of 9.4(3) MHz.
By deconvolving this Lorentzian contribution from the measured profile
and assuming temporal Gaussian laser pulses leading to a $\sqrt{2}$ broadening in each frequency-doubling step,
we estimate  7.2(4) MHz for the bandwidth of the Ti:Sa laser for chirp compensated pulses. However, this is a slight over-estimate since the broadening effect of temporal overlap by the ionization pulse in the H$_2$ experiment is not accounted for.
The difference in line centers of the two curves in Fig.~\ref{fig:AllanDev}b, is largely due to cw-pulse frequency offset compensation but also includes contributions from ac-Stark effects in the spectroscopic measurements~\cite{Cheng2018}.

Figure~\ref{fig:AllanDev}a shows that the Allan deviation of the Ti:Sa output minimizes at  20 kHz (at 10 s), corresponding to a relative uncertainty of $5 \times 10^{-11}$.
This can in principle be considered as accurate, since the chirp measurements were performed with respect to a cw laser locked to a Cs-frequency standard with an accuracy of $\sim 10^{-12}$.
In addition, we performed stability measurements of the pulsed output referencing  against a molecular resonance [H$_2$, X-GK Q(1) line in two-photon excitation~\cite{Cheng2018}], where each data point was averaged over 50 laser shots (at 10 Hz operation).
It is noted that in addition to the active frequency control, the residual chirp was measured for every laser shot for (out-of-loop) numerical compensation of the frequency.
This assessment of the chirp yields an upper limit  to the absolute accuracy of $1.5 \times 10^{-10}$.
The latter result improves a previous experiments with an intra-cavity EOM~\cite{Hori2009}, by a factor of five, and the passively compensated Ti:Sa oscillator-amplifier system~\cite{Hannemann2007b} by a factor of 15.

\section*{Funding Information}

The work was funded by an Advanced grant (670168) of the European Research Council and by the
National Natural Science Foundation of China (21688102) and Chinese Academy of Sciences (XDB21020100).

\section*{Disclosures}
The authors declare no conflicts of interest.



\end{document}